\documentclass[12pt,preprint]{aastex} %%%%one-column, single-spaced document

\usepackage{url}
\usepackage{natbib}
\usepackage{booktabs}
\usepackage{graphicx,color}
\newcommand{\degph}{\mbox{\arcdeg h$^{\rm -1}$}}
\newcommand{\kob}{{\it Kodaikanal Observ. Bull.}}
\shorttitle{Rotating Sunspots and Eruptive Activity}
\shortauthors{Vemareddy et al}

\begin{document}

\title{On the Role of Rotating Sunspots in the Activity of
Solar Active Region NOAA 11158}

\author{P.~ Vemareddy$^1$, A.~ Ambastha$^1$, and R.~ A.~ Maurya$^2$}
\affil{$^1$Udaipur Solar Observatory, Physical Research Laboratory,
Udaipur-313 001,India.} \affil{$^2$Astronomy Program, Department of
Physics and Astronomy, Seoul National University, Seoul 151-747,
Korea}
\email{vema@prl.res.in, ambastha@prl.res.in, ramajor@astro.snu.ac.kr}

%%%%%%%%%%%%%%%%%%%%%%%%%%%%%%%%%%%%%%%%%%%%%%%%%%%%%%%%%%%%%%%%%%%%%%%%%%
%% Abstract %
%%%%%%%%%%%%%%%%%%%%%%%%%%%%%%%%%%%%%%%%%%%%%%%%%%%%%%%%%%%%%%%%%%%%%%%%%%

\begin{abstract}
We study the role of rotating sunspots in relation to the evolution of various 
physical parameters characterizing the non-potentiality of the active region NOAA 11158 
and its eruptive events using the magnetic field data from the Helioseismic and 
Magnetic Imager (HMI) and multi-wavelength observations from the Atmospheric Imaging 
Assembly (AIA) on board Solar Dynamics Observatory (SDO). From the evolutionary study of 
HMI intensity and AIA channels, it is observed that the AR consists of two major rotating 
sunspots one connected to flare-prone region and another with CME. The constructed space-time 
intensity maps reveal that the sunspots exhibited peak rotation rates coinciding with the 
occurrence of the major eruptive events. Further, temporal profiles of twist parameters, viz., 
average shear angle, $\alpha_{\rm av}$, $\alpha_{\rm best}$, derived from HMI vector magnetograms 
and the rate of helicity injection, obtained from the horizontal flux motions of HMI line-of-sight 
magnetograms, corresponded well with the rotational profile of the sunspot in CME-prone 
region, giving predominant evidence of rotational motion to cause magnetic non-potentiality. 
Moreover, mean value of free-energy from the Virial theorem calculated at the photospheric 
level shows clear step down decrease at the on set time of the flares revealing unambiguous 
evidence of energy release, intermittently that is stored by flux emergence and/or motions 
in pre-flare phases. Additionally, distribution of helicity injection is homogeneous in CME prone
region while it is not and often changes sign in flare-prone region. This study provides 
clear picture that both proper and rotational motions of the observed fluxes played significant 
role to enhance the magnetic non-potentiality of the AR by injecting helicity, twisting the 
magnetic fields thereby increasing the free energy, leading to favorable conditions for 
the observed transient activity.\end{abstract}


%%%%%%%%%%%%%%%%%%%%%%%%%%%%%%%%%%%%%%%%%%%%%%%%%%%%%%%%%%%%%%%%%%%%%%%%%%
%% 1. Introduction %
%%%%%%%%%%%%%%%%%%%%%%%%%%%%%%%%%%%%%%%%%%%%%%%%%%%%%%%%%%%%%%%%%%%%%%%%%%
\section{Introduction}
\label{Intro}

Activity level of the Sun is defined by eruptive events like coronal
mass ejections (CMEs) and flares occurring in the ARs consisting of
groups of sunspots having positive and negative magnetic polarities.
During the evolution of ARs, the sunspots exhibit motions, both
proper and rotational, that lead to the storage of energy in the AR
by increasing its magnetic non-potentiality. The eventual release of
stored energy then occurs in the form of the eruptive events, such
as, flares and CMEs. An investigation of the characteristics of
sunspots can thus help towards the understanding of transient solar
activities.

Rotational motions of and around sunspots have been observed by
several authors in the past
\citep{evershed1910,batnagar1967,mcintosh1981}. More recent studies
have also reported sunspot rotations about umbral center as well as
about another sunspot with the availability of high spatial and
temporal resolution data from space borne as well as ground based
observatories \citep{brown2003,zhang2007,yan2009}. Rotational
motions can shear and twist the overlying magnetic field structures
thereby storing excess energy. Vertical current, current helicity
and transport helicity are some of the quantitative parameters
derived from observations that are related to the twisting of
fieldlines anchored in the moving footpoints. CMEs that disturb the
space environment are thought to occur due to an over-accumulation
of helicity and its ejection from the corona to the interplanetary
space. Recently, \cite{zhang2008} reported that helicity injection
as inferred from the rotational motion of sunspots can be comparable
to that as deduced for the entire AR by local correlation tracking
(LCT) method. They also found good spatial and temporal correlation
of sunspot rotational motion with two homologous flares.

%The problems of the origin of rotational motion of sunspots and its
%mechanism are still not well understood. \citet{knoska1975}
%suggested that sunspot rotation is related to the dynamics of the
%developing magnetic field of the AR. Coriolis force and differential
%rotation are often thought to be the origin of the sunspot rotation
%(Ding, Hong, and Wang, 1987). \citet{zhao2003} found evidence of
%structural twist beneath the visible surface of a rotating sunspot
%by using time-distance helioseismology. Chae et al. (2003) proposed
%that rotational motion of a sunspot can be driven by the expansion
%of the coronal segment of a twisted flux tube. Further, sheared
%layers in the depth range of 0-10 Mm have been discovered in a large
%sample of ARs \citep{maurya2010b}. These structures imply the
%presence of complex flow structures beneath the photosphere which
%may be responsible for the observed rotational motion.

The consequences of sunspot rotations and related changes, both in
photosphere and corona, have been investigated by several workers.
Relationships between the sunspot rotation and coronal consequences
\citep{brown2003,tian2006,tian2008}, flare productivity
\citep{yan2008b,zhang2008}, statistics of the direction of rotating
sunspots, helicities \citep[and references there in]{yan2008b}, and
the association of flares with abnormal rotation rates
\citep{hiremath2005} are some studies in this direction.

In this study, we report on the rotating sunspots observed during
February 11-16, 2011, in AR NOAA 11158. We present the results
related to the changes in various parameters, viz., vertical
current, twist parameters, helicity injection, and the free energy of
the AR during its evolution. Further, we examine whether the
rotating sunspots were associated to the eruptive flare and CME
events of the AR. Our aim is towards finding association of the
observed uncommon rotation of the sunspots of this AR in terms of 
changes in various physical parameters. We have organized the paper
as follows: Observational data and analysis procedure is described in 
Section~\ref{Data} which is followed by the results with possible discussion in
Section~\ref{Res}. Finally, the summary of the study
is presented in Section~\ref{SumConc}.
%%%%%%%%%%%%%%%%%%%%%%%%%%%%%%%%%%%%%%%%%%%%%%%%%%%%%%
%%                OBSERVATIONS                      %%
%%%%%%%%%%%%%%%%%%%%%%%%%%%%%%%%%%%%%%%%%%%%%%%%%%%%%%
\section{Observational Data and Analysis Procedure}
\label{Data} The observational data used in this study are obtained from the
Helioseismic and Magnetic Imager (HMI; \citealt{schou2012}) and the Atmospheric
Imaging Assembly (AIA, \citealt{lemen2012}) on board {\it Solar Dynamics
Observatory} (SDO).

AIA observes the sun in ten different wavelengths in soft X-ray, EUV
and UV regimes to study the processes occurring at various
atmospheric layers {\it i.e.}, chromosphere, transition region and
corona, at a pixel size of 0.6\arcsec~and 12 second cadence. In
addition, we also obtained high resolution G-band 4300\AA~and Ca {\sc ii}
H 3970\AA~images from Solar Optical Telescope (SOT;
\citealt{tsuneta2008}) on board Hinode for understanding the
associated chromospheric structures. Data on the CME and flare
initiation times were obtained from the
websites\footnote{\url{http://spaceweather.gmu.edu/seeds/}},
\footnote{\url{http://www.solarmonitor.org}}.

HMI observes the full solar disk in the Fe {\sc i} 6173\AA~ spectral
line with a spatial resolution of 0.5\arcsec~/pixel and provides four
main types of data at a cadence of 12 minutes: dopplergrams (maps of
solar surface velocity), continuum filtergrams (broad-wavelength
maps of the solar photosphere), and both line-of-sight (LOS) and
vector magnetograms. Filtergrams are obtained at six wavelength
positions centered at the Fe {\sc i} line to compute Stokes
parameters I, Q, U, V. These are then reduced with the HMI science
data processing pipeline to retrieve the vector magnetic field using
the Very Fast Inversion of the Stokes Vector algorithm (VFISV;
\citealt{borrero2011}) based on the Milne-Eddington atmospheric
model. The inherent 180\arcdeg~ azimuthal ambiguity is resolved
using the minimum energy  method \citep{metcalf1994,leka2009}. Data
processing is carried out by the HMI vector team
\citep{hoeksema2012} and the products are made available to the
solar community.

We used the high resolution continuum images of NOAA 11158 obtained from the
HMI for the rotation  measurement of sunspots. In order to estimate the
twisting in various sites of the AR, we used vector magnetograms projected and
remapped to heliographic coordinates. The intensity images obtained from AIA
were used to examine the coronal changes associated to the photospheric
motions. All images both from AIA and HMI taken at different times were aligned
by differentially rotating to an image at central meridian by non-linear
remapping method \citep{howard1990}. For our analysis, we have used the
standard SolarSoft routines (SSW; \citealt{freeland1998}) implemented in
Interactive Data Language(IDL).

In order to further characterize the association of rotational
motion on non-potentiality and in turn on the observed transient
activity of the AR, we examine the evolution of magnetic fluxes and
the derived parameters (see \citet{leka2003a} for definitions of various non-potential 
parameters) therefrom. We evaluate the vertical currents
and shear or twist parameters to estimate the extent of non-potentiality. The
complexity developed through the flux motions by helicity injection
rate and consequent magnetic energy storage are also analyzed.
Procedures for these calculations are briefly outlined below:

\begin{enumerate}

\item We evaluate the total absolute flux in sub-regions R1 and R2 by summing over
all the pixels $F=\sum |B_z|$. The vertical current is computed from
the Ampere's law $\nabla\times\bf{B}=\mu_0\bf{J}$, where the
permeability of free space $\mu_0=4\pi\times10^{-7}$ Henry/meter.
Using the z-component of this relation, one can deduce the total 
absolute vertical current as $J_z=\sum |(\nabla \times \mathbf{B})_z/\mu_0|$.

\item \emph{Magnetic Shear}: Non-potential nature of magnetic field is 
characterized by magnetic shear defined as difference between directions of 
observed transverse field and potential transverse field 
\citep{ambastha1993,wangh1994} and is given by
\begin{equation}
S = B\cdot\theta=B\cdot\left(\cos^{-1}\frac{\mathbf{B}\cdot\mathbf{B_p}}{BB_p}\right)\label{Eq_ShD}
\end{equation}
where $B=|\mathbf{B}|$ is magnitude of observed magnetic field and $B_p=|\mathbf{B_p}|$ 
is magnitude of potential field computed from fourier method \citep{gary1989}. 
The weighted shear angle in a region of interest with N pixels can then be obtained as
\begin{equation}
{\rm WSA} = \frac{\sum_i^{N}S_i}{\sum_i^N B_i}\label{Eq_WSA}
\end{equation}
and will estimate the overall angle of departure from potential field configuration.

\item \emph{Alpha best ($\alpha_{\rm best}$)}: The force free parameter $\alpha_{\rm best}$
is used as a proxy to monitor the extent of twist of the magnetic
fieldlines in an AR. For this parameter, the extrapolated force free
fields best fit the observed transverse fields in the sense of a
minimum least-squared difference on the photosphere. The procedure
of the fitting method for calculating $\alpha_{\rm best}$
\citep{pevtsov1994,hagino2004} is as follows. For a given $\alpha$
and with observed $B_z$, we compute linear-force free field whose
transverse component is $B_{t,cal}$. We then search for the $\alpha$
that gives minimum residual ($R(\alpha)$) of transverse component in
a range of values through the following equation
\begin{eqnarray}
R{\alpha} = \frac{\sum[B_{t,obs}(x,y)-fB_{t,cal}(x,y,\alpha)]^2}{\sum
B^2_{\rm t,obs}} \label{Eq_Resid_alp} \\
{\rm where} \hspace{0.2in}  f=\frac{\sum B^2_{\rm t,obs}}{\sum B^2_{\rm
t,cal}(x,y)} \
\end{eqnarray}

The error in the fitting of residual and $\alpha$ that is effected in
estimating $\alpha_{\rm best}$ is given by
\begin{equation}
\delta\alpha^2_{best}=\frac{2\delta B_t \sqrt{R_0}}{R'\sqrt{\sum B^2_{\rm
t,obs}}}\label{Eq_Err_Alpb}
\end{equation}

\item \emph{Average Alpha ($\alpha_{\rm av}$)}: Another proxy for representing the
twist is obtained by taking the average of $\alpha$ from the
force-free assumption of photospheric fields, i.e.,
$\nabla\times\textbf{B}=\alpha\textbf{B}$. Following
\cite{pevtsov1994} and \cite{hagino2004}, it is given by:

\begin{equation}
\alpha_{\rm av} = \frac{\sum J_z(x,y){\rm sign}[B_z(x,y)]}{\sum |B_z|}
\label{Eq_Alp_av}
\end{equation}
The error in $\alpha_{\rm av}$ is deduced from the least squared
regression in the plot of $B_z$ and $J_z$, and is given by:
\begin{equation}
\delta\alpha_{\rm av}^2 = \frac{\sum[J_z(x,y)-\alpha_{\rm
av}B_{z}(x,y)]^2/|B_{z}(x,y)|}{(N-1)\sum|B_{z}(x,y)|} \label{Eq_Err_Alpav}
\end{equation}
where N is the number of pixels where $|B_t|\geq150$G (with 150G
assumed as the lower cut-off for the transverse field.)

\item \emph{Helicity Injection}:
One uses the helicity injection rate to quantify the twist due to
photospheric flux motions. Following \cite{pariat2005}:
\begin{equation}
\frac{dH}{dt}=
\frac{-1}{2\pi}\int_{S}\int_{S'}\frac{[(\textbf{x}-\textbf{x}')\times(\textbf{u}-\textbf{u}')]_n}{|\textbf{x}-\textbf{x}'|^2}B'_n(\textbf{x}')
B_n(\bf{x}) dS'dS \label{Eq_par} \\
\end{equation}
where $\textbf{u}(\textbf{u}')$ is the foot-point velocity at
position vector $\bf{x}(\bf{x}')$, and $B_n$ is the normal component
of magnetic field from observations. This equation shows that the
helicity injection rate can be understood as the summation of
rotation rates of all the pairs of elementary fluxes weighted with
their magnetic flux. Integration of above equation in the observed
time interval provides the accumulated helicity, $\Delta H=\int
(dH/dt)\Delta{t}$ that is injected due to foot-point shear motions.

\item \emph{Magnetic Free Energy}: One can get the estimate of non-potentiality
by the magnetic virial theorem
\citep{chandrasekhar1961,molodensky1974,low1982}, which gives the
total force-free magnetic energy as
\begin{equation}
E_m=\int_{z>0}\frac{\bf{B}^2}{8\pi}dV=\frac{1}{4\pi}\int_{z=0}(xB_x+yB_y)B_z
dxdy \label{Eq_FreeEne}
\end{equation}
where $B_x$, $B_y$ are the horizontal and $B_z$ the vertical
components of magnetic field $\bf B$ at position (x, y) on the solar
surface z=0. The applicability of this equation to the realistic
situations like solar photospheric boundary must be restricted by
force-freeness of the field and flux balance conditions. Despite its
applicability, it should have some useful information about energy
storage and release. Since the potential field state is a minimum
energy state, by subtracting the potential field energy from the
force-free energy one gets the upper limit of the free-energy
available in the AR to account for the flares and CMEs.

However, the photospheric field is not strictly force-free
\citep{metcalf1995,gary2001} and flux balance in the region of
interest are not fully satisfied. Therefore, one obtains incorrect 
estimates of the free-energy. To derive the error estimate for the force-free 
energy measurement, we have adopted the Pseudo-Monte Carlo method \citep{metcalf2005},
which gives the mean and variance of energy estimate by changing the
origin of coordinate system in all possible ways. It includes both
the statistical error as well as the error due to the departure from
the force-free state of the fields.

\end{enumerate}

We have examined the errors given with the vector field data after
inversion of stokes profiles. The maximum error in transverse field
$B_t$ in the entire data set varied in the range of 20-35G. The
total field strength was observed to vary by 50G over a day due to
various factors \citep{hoeksema2012}. Therefore, we carried out the
calculations with $|B_{t}|\ge150$G and $|B_z|\ge15$G and neglected
the pixels having magnetic fields below these thresholds, simultaneously.

As require by Equation~\ref{Eq_par}, the horizontal motions of photospheric fluxes 
using LOS magnetic field data obtained at a cadence of 12 minutes by the 
method of Affine Velocity Estimator (DAVE; \citealt{schuck2006}). It is a local 
optical flow method that determines the mass velocities within the windowed region. 
Further, it adopts an affine velocity profile specifying velocity field in the 
windowed region about a point and constrains that profile to satisfy the induction equation.
The maximum rms velocities of flux motions in the entire AR were found to be distributed in
the range 0.6-0.9 kms$^{-1}$. These velocity maps of mass motions revealed
strong shear flows that included both proper and rotational motions. With the velocities 
of these flux motions, we computed the helicity flux density maps of the whole AR at pixels 
above 10G to reduce computation time, then helicity injection rate, and accumulated helicity 
using Equation~\ref{Eq_par}.
 
%However,along the penumbral fibrils of SP3, only the continuous outward moat flows were 
%noticed in the studied time period.

%%%%%%%%%%%%%%%%%%%%%%%%%%%%%%%%%%%%%%%%%%
%%           RESULTS                    %%
%%%%%%%%%%%%%%%%%%%%%%%%%%%%%%%%%%%%%%%%%%
\section{Results and Discussion}
\label{Res}
The AR NOAA 11158 was a newly emerging region that first appeared as
small pores on the solar disk on February 11, 2011 at the
heliographic location E33S19. Subsequently it grew rapidly by
merging of small pores to form bigger sunspots and developed to
$\beta\gamma\delta-$ magnetic complexity on 2011 February 13. The HMI
intensity maps in Figure~\ref{IntMos} show the spatio-temporal
evolution of AR 11158 during 2011 February 11-16. The prominent
negative and positive polarity sunspots are labeled in frame (d) as
SN1--SN3, SP1--SP3, respectively, along with the overlaid LOS
magnetic field contours in frame (e) depicting the polarity
distribution of the AR. The overall configuration of the AR is seen
to be of quadrupolar nature. The proper motion of individual
sunspots during February 13--16 are marked by arrowed curves traced by
centroids of the sunspots as shown in frame (f).

In what follows, we shall present evolution of the AR during 2011
February 13-15 using coronal observations in different wavelengths
along with the changes in its magnetic and velocity field
structures, the related physical parameters and the flare/CME
activities.

\subsection{Evolution and the Activity of the AR}
A careful examination of the animations of magnetogram
and intensity maps revealed significant counter clock-wise (CCW)
rotation of SN1 and clock-wise (CW) rotation of SP2 during February
13-15. We selected two sub-regions based on distinctive flux motions
and associated transient activities. These are sub-region R1
covering SN1, and R2 around SN2, SN3, and SP2 as shown in the
rectangular boxes in Figure~\ref{IntMos}(d).

The spatial evolution in R2 displayed a large shearing motion of SP2
that itself rotated clock-wise during February 13-15. It then
detached from SN2 and moved towards SP3 with small patches of both
polarities appearing and disappearing over short periods of time.In R1, a
small positive polarity site SP1 emerged to the north of SN1 that
rotated in a similar counter-clockwise direction along with a proper
motion away from SN1. The rotation of SN1 and its associated SP1
twisted the field lines. These motions appears to have resulted in a
large magnetic non-potentiality in this sub-region. This followed the first 
X-class event of the current solar cycle 24, an X2.2 flare in R2 on February 15. 
In addition, several flares of smaller magnitude, including four
M-class and several C-class occurred during this period; all 
associated to R2 except the M2.2 flare of 14 February which is observed to 
originate from R1. Many intermittent CMEs were also launched from this AR. These
flares and CMEs (listed in Table~\ref{TabFlCm}) appear to be related
to the observed motions in the AR. We have manually traced the the
QuickLook AIA images to look for mass ejections in 304\AA~channel
locating the sub-region they belonged to and found that most of the
CMEs originated from R1 (except the one associated with the X2.2
flare from R2). This inference is further confirmed by the STEREO
space-craft information\footnote{\url{http://spaceweather.gmu.edu/seeds/}}.

\begin{table*}[h]
\centering \caption{List of Flares and CMEs}
\begin{tabular}{l l l l}
  \hline                                                                                 
  AR       &  Date      &  Flares                             &   CMEs                \\
  (NOAA)   &dd/mm/yyyy        & magnitude(time UT)& (time UT)                          \\
  \hline                                                                                   
  11158    & 13/02/2011 & C1.1(12:36),C4.7(13:44),M6.6(17:28) & 21:30,23:30            \\
           & 14/02/2011 & C1.6(02:35),C8.3(04:29),C6.6(06:51) & 02:40,07:00,12:50      \\
           &            & C1.8(08:38),C1.7(11:51),C9.4(12:41) & 17:30,19:20            \\
           &            & C7.0(13:47),M2.2(17:20),C6.6(19:23) &                         \\
           &            & C1.2(23:14),C2.7(23:40)             &                         \\
           & 15/02/2011 & C2.7(00:31),X2.2(01:44),C4.8(04:27) & 00:40,02:00,03:00      \\
           &            & C1.0(10:02),C4.8(14:32),C1.7(18:07) & 04:30,05:00,09:00      \\
           &            & C6.6(19:30),C1.3(22:49)             &                        \\
  \hline                                                                               \\
\end{tabular}                                                                           \\
\label{TabFlCm}
\end{table*}

\begin{figure}
  \centering
  \includegraphics[width=.99\textwidth,clip=,bb=73 370 409 651]{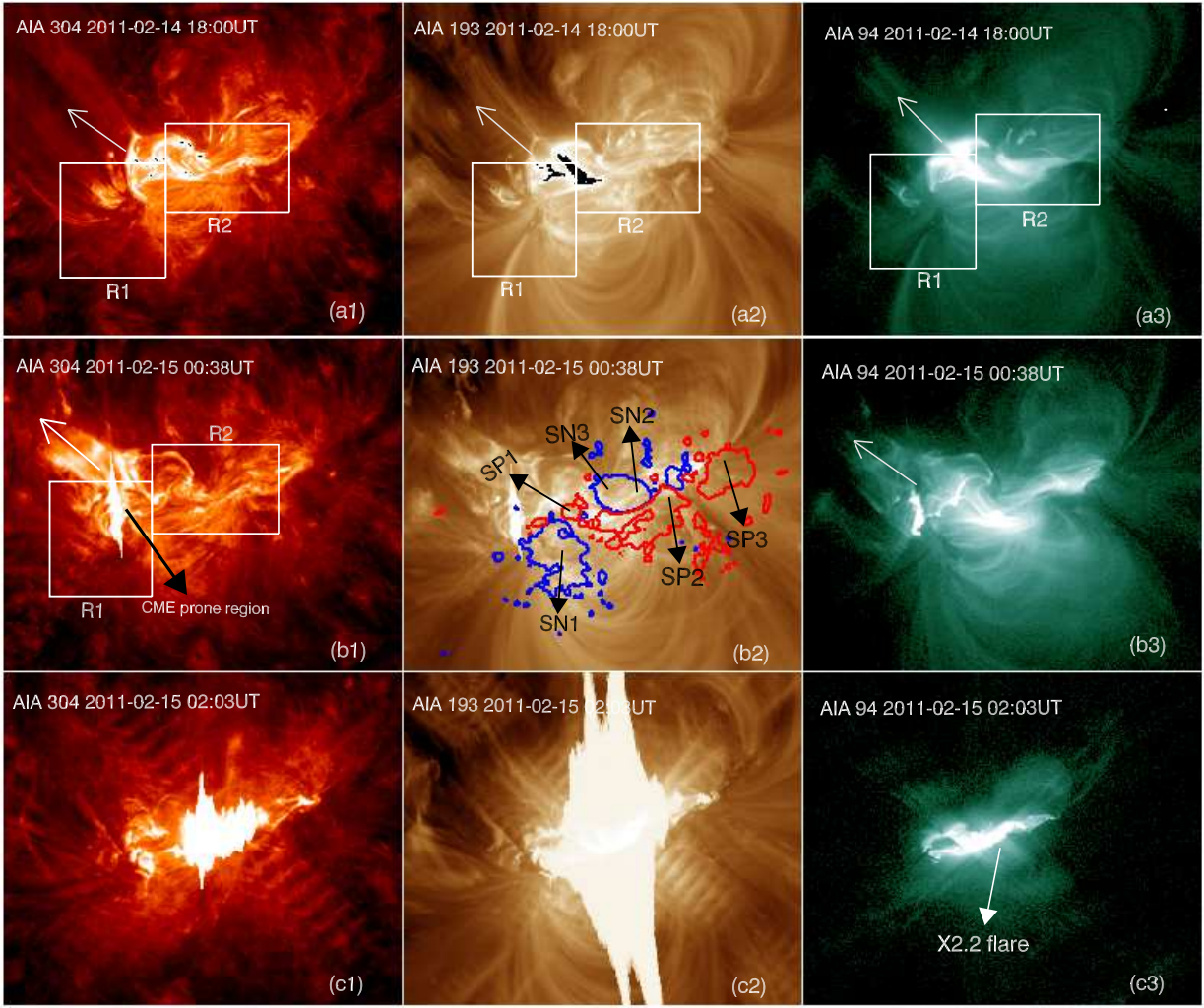}
   \caption{The coronal activity observed during the evolution of the AR in AIA \
   wavelengths. (a1--a3): A mass expulsion (dashed arrow) from R1 which eventually 
   turned to a CME on 2011 Feb 14/18:00UT. (b1--b3): Another mass expulsion 
   observed on 2011 Feb 15/00:36 UT from the same location. (c1): Bright emission 
   seen during the X2.2 flare, saturating the 193\AA~detector (c2) and the 
   twisted flux ropes along the polarity inversion line as seen in 94\AA~ (c3).}\label{aia_plot}
\end{figure}

Morphological changes observed in chromosphere, transition region,
and coronal regions corresponding to some of the transient events
are shown in Figure~\ref{aia_plot} using AIA observations in 304,
193 and 94\AA~wavelengths. We further examine the characteristics
of non-potentiality in the selected sub-regions.

Figure~\ref{aia_plot} (a1--a3) shows a large mass expulsion from
sub-region R1 that turned into a large CME on February 14/18:00UT with the 
associated flare M2.2 at 17:20UT. It appeared to occur when the rotation of 
SN1 had attained the maximum speed. This
CME began at 17:30UT and continued well past 19:30UT as observed
from the AIA/304\AA~movies. The ejected plasma (marked by arrow in a1) is 
visible in 304\AA~as it is sensitive to the chromospheric temperature 
($\approx 10^4$K). However, it is only partially visible in 193\AA~and 
not discernible in 94\AA~except for the cusp shaped fieldlines filled 
with the hot bright plasma at $\sim10^6$K. Another such mass ejection 
of February 15/00:36UT that ensued from the same location is shown in 
Figure~\ref{aia_plot}(b1--b3). The LOS magnetic field contours are overlaid 
in frame b2 for reference, depicting the fieldline connectivities from 
the magnetic footpoints in the labeled sunspots. From the animations of magnetic 
and intensity images and coronal fieldline connectivity, a significant role of 
new emerging flux at SP1 can be inferred in these expulsions.

Figure~\ref{aia_plot} (c1--c3) shows the intense X2.2 flare which
occurred on February 15 in sub-region R2. This event was also
associated with a halo-CME which began at 01:44UT. This flare
produced abnormal magnetic polarity reversal and Doppler velocity
enhancement during the flare's impulsive phase \citep{maurya2012}.
As mentioned earlier, the rotating flux footpoints, i.e.,
the sunspots, twisted the overlying magnetic structures in opposite
directions. The rotational motion of the positive polarity spot SP2
about its negative polarity counterpart SN2 developed strongly
twisted fieldlines along the polarity inversion line (PIL), clearly
visible as a ``sigmoidal'' structure (c3). As reported earlier, such
sigmoidal magnetic field structures are more likely to erupt
\citep{canfield1999}.

\subsection{Measurement of Sunspot Rotation}
To derive the rotational parameters of the sunspots, we followed the 
procedure as explained in \citet{brown2003}. Essentially, we pursued
the motion of a distinct feature in the penumbra as it is
comparatively richer than the umbra. We uncurl the annular region of
penumbra by transforming from the Cartesian (x--y) frame to the
polar (r-$\theta$) plane. In this way, we can measure the position
angle of the observed feature as displaced from its previous
position as the sunspot rotated. We identified the centroid of the
sunspot SN1 by setting the 50\% contrast between the umbra and
penumbra in the time sequence of intensity images around the region
of a circular disk of radius 18 arc-sec. This circular region was
then remapped on to the polar coordinate system from the Cartesian,
so that annular features appeared as straight strips from umbra to
penumbra. Once this is done, any portion of the strip could be
tracked in time to determine the angular distance by which a
particular feature is displaced.

\begin{figure}
  \centering
  \includegraphics[width=.95\textwidth,clip=,bb=8 4 424 174]{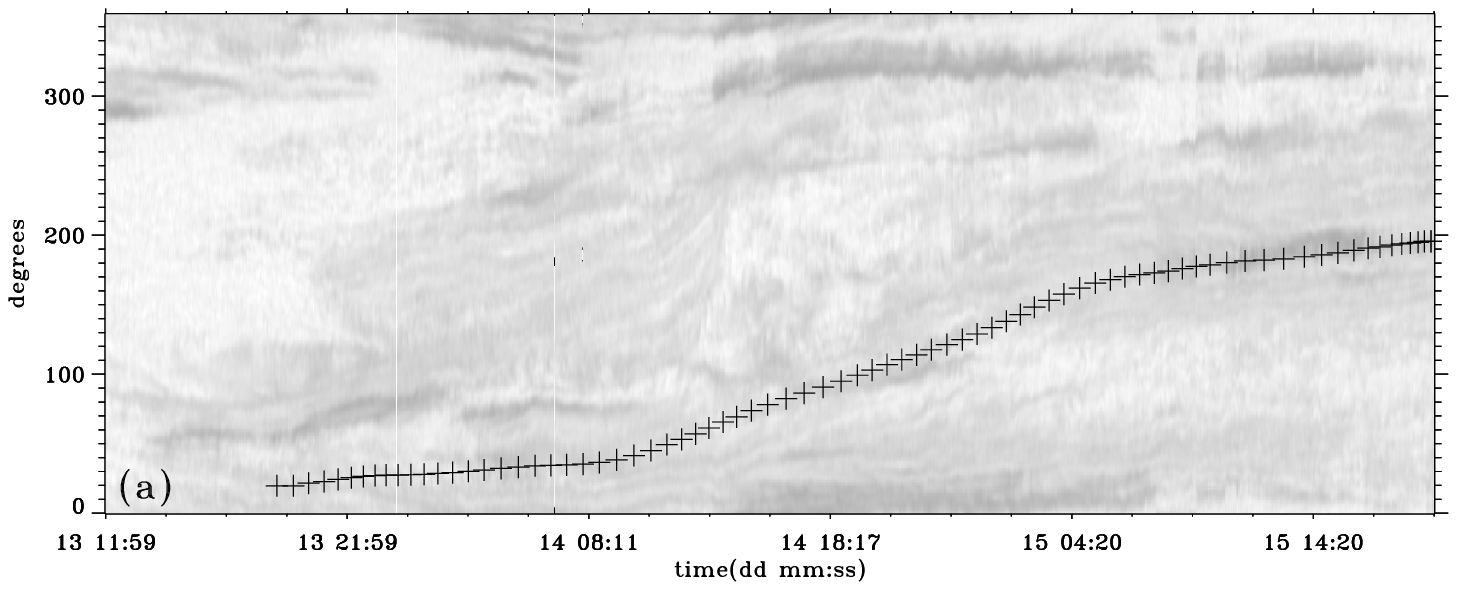}
  \includegraphics[width=.95\textwidth,clip=,bb=8 4 424 174]{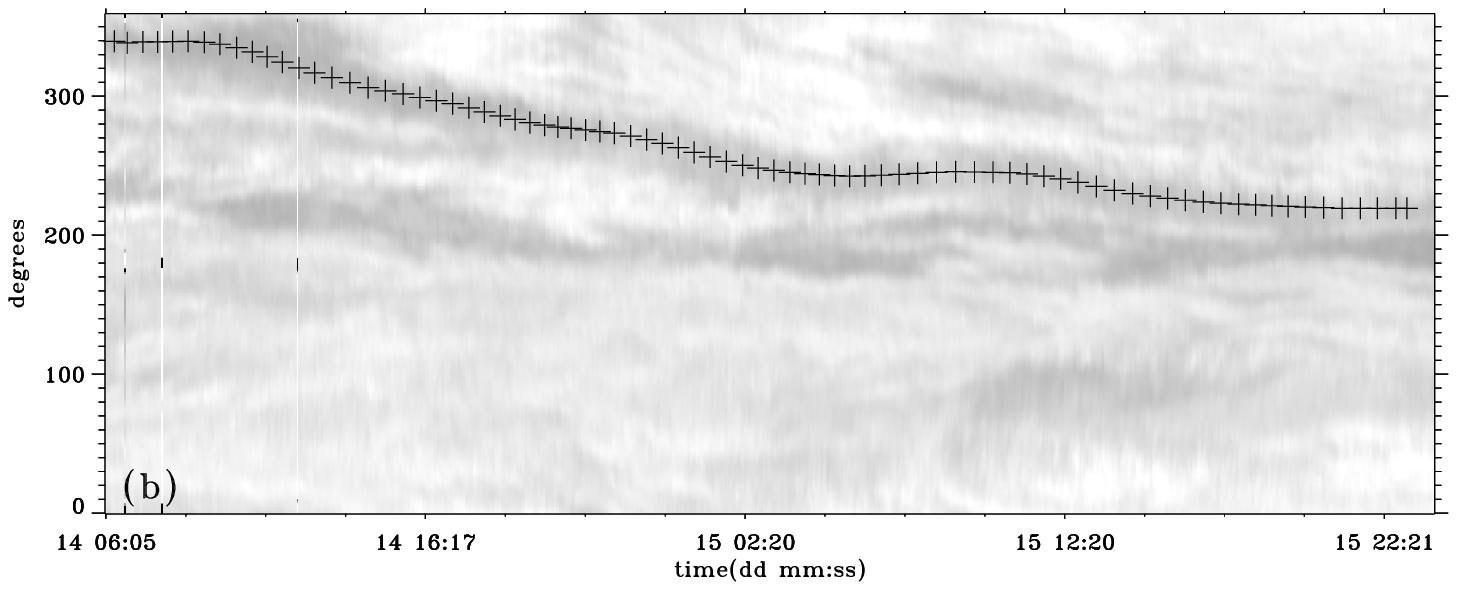}
  \includegraphics[width=.48\textwidth,clip=,bb=14 4 298 210]{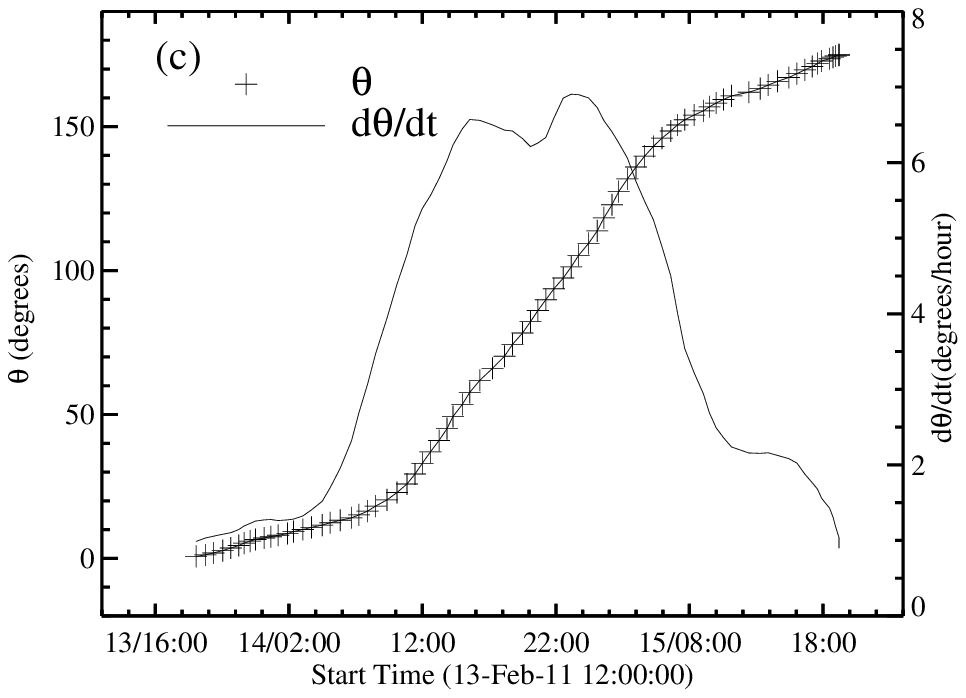}
  \includegraphics[width=.48\textwidth,clip=,bb=14 4 298 210]{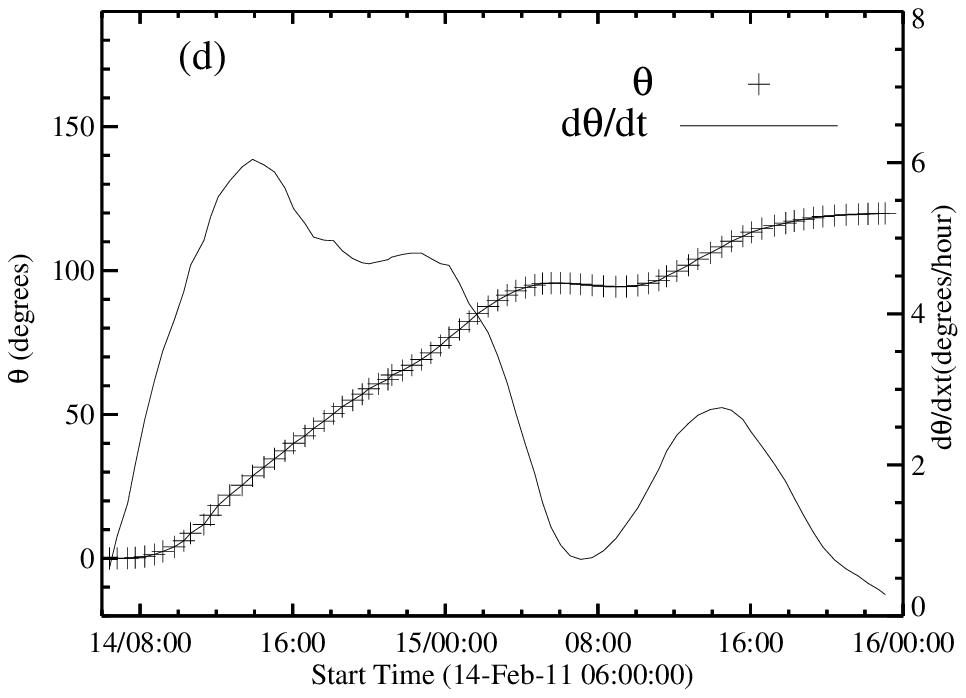}
  \caption{Space-time map of a strip in (a) the sunspot SN1 (at a distance of 7 
  arc-sec from the sunspot's centroid), (b) sunspot SP2 (at a distance of 11 arc-sec 
  from the sunspot's centroid), after remapping the annular region of the sunspot 
  to the r-$\theta$ plane. Rotation angle ($\theta$) and rotation rate 
  ($d\theta/dt$) of the sunspot's umbrae are inferred from the well-appearing 
  diagonal feature in (a)-(b) sampled along the path marked by ``+'', and plotted with time 
  in (c)-(d), respectively.}\label{FigRot}
\end{figure}

We constructed the space-time diagrams for both the rotating
sunspots, SN1 and SP2, extracted at the radii of 11\arcsec~and
7\arcsec~from their umbral centroid, as shown in
Fiugre~\ref{FigRot}(a-b), respectively. Rotational motion of the features 
in the penumbral regions can be seen as diagonal bright or dark streaks.
The uncurling starts at westward point chord connecting the centroid
position and the outer circle of annular portion and proceeds
anti-clockwise about the sunspot centroid. These diagrams clearly
show the counter-clock wise (CCW) rotation of SN1 in sub-region R1,
and the clock wise (CW) rotation of SP2 in sub-region R2. A well
observed streak is traced and plotted as marked by ``+" at the
respective positions as angular displacement ($\theta$) and its time
derivative i.e., angular velocity ($d\theta/dt$) with time in
Figure~\ref{FigRot}(c and d).

The sunspot SN1 started rotating early on from February 14 which
continued till February 15/18:00UT. During the 24h period of
February 14, it rotated by over 100\arcdeg~at an average rotation
rate of 4\degph~, peaking at 7\degph~at around 23:50UT, i.e., just
3h before the X-flare. The timings of mass expulsions from this
sub-region (cf., Figure~\ref{aia_plot}(a1)) also coincided well with
the fast rotation rate of SN1. Except for the CME associated with
the X2.2 flare, all other CMEs listed on February 15 were found to
be associated with the sub-region R1 even in the decreasing phase of
its rotation rate. This suggests that the new emerging flux SP1 and 
large rotation rate of SN1 might have played a role in the process 
of sigmoidal structure formation which gave rise to the CMEs that 
were often associated with flares.

A similar analysis was carried out for evaluating the rotation of
SP2 in order to examine its role in the X2.2 flare as reported by
\cite{jiang2012}. For the sake of completeness and consistency, 
we have reproduced the results for SP2 in Figure~\ref{FigRot}. 
It exhibited rotation rates of 4.48\degph~on February 14 before the X2.2 
flare and 1.92\degph~on February 15. These are consistent with the results 
deduced by \cite{jiang2012}. Moreover, it rotated by 95\arcdeg~till 
February 15/04:00 UT which is also within the error limits of our manually 
tracked feature to their reported value of 107\arcdeg. Notably, a small 
rotation rate of 0.9\degph~was observed after the flare from the time profile
showing that it rotated by a total of 25\arcdeg~till the end of the
day. As is evident, there existed proper motion in addition to the
rotational motion of SP2 (Figure~\ref{IntMos}(f)). Therefore, it may
not be appropriate to attribute the rotational motion alone to the
X2.2 flare and other events observed with this region.

It is thus evident that the two sub-regions consisting of the
sunspots with large rotation rates were essentially the sites from
where flares and CMEs originated in NOAA 11158. We speculate that
the rotational motion led to the adequate storage of energy and injection of
helicity which subsequently played the predominant role in the
eruptions. In what follows, we shall discuss these aspects in
further detail using the physical parameters deduced from the velocity and
magnetic field measurements.

\subsection{Evolution of Physical parameters in sub-regions R1 and R2}

We now present a detailed description of the magnetic and velocity
field observations and their derived parameters, as discussed in 
Section~\ref{Data}, in the sub-regions R1 and R2 containing the rotating 
sunspots.

\begin{figure}
  \centering
  \includegraphics[width=.60\textwidth,clip=,bb=66 41 487 773]{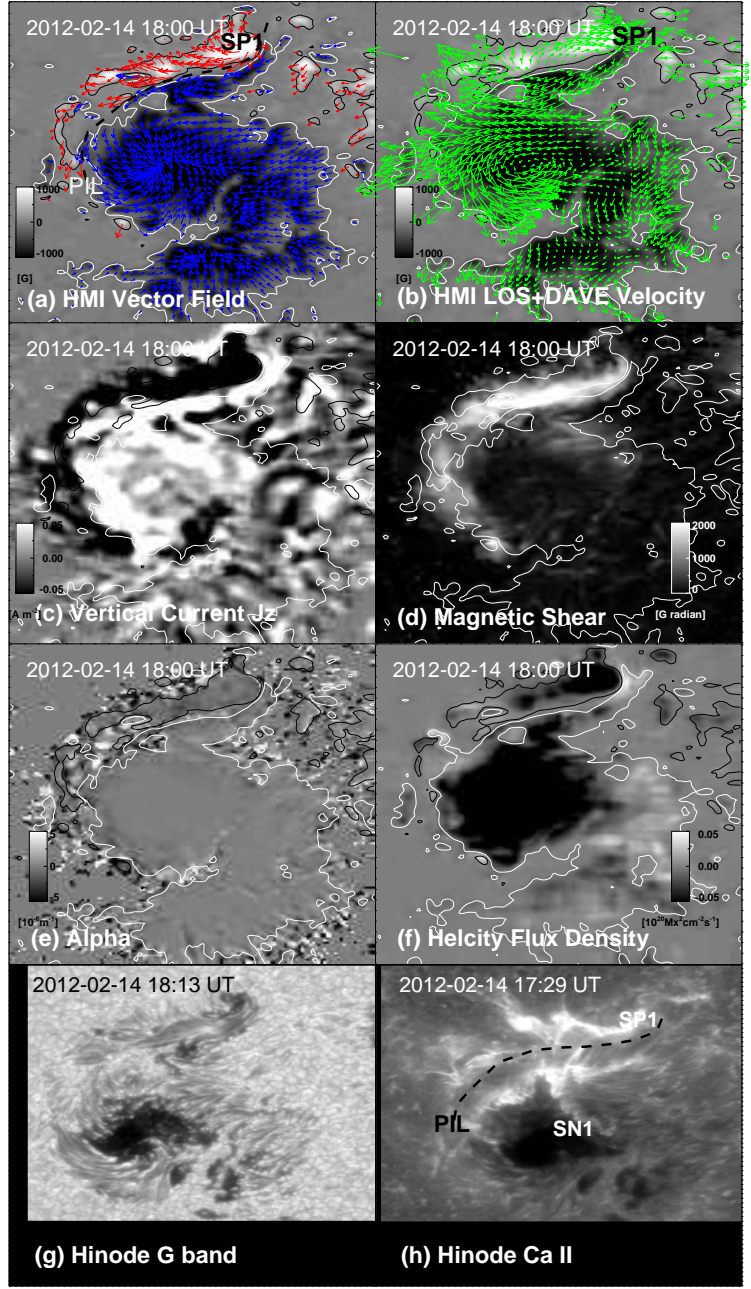}
   \caption{Sub-region R1 on February 14/18:00UT: (a) transverse magnetic field 
   ($B_t$) vectors overlaid on LOS magnetic field drawn with polarity inversion 
   line (dashed curve), (b) horizontal velocity vectors showing vortical-like pattern 
   due to rotation of SN1, (c) vertical current ($J_z = (\nabla\times\mathbf{B})_z/\mu_0$) 
   distribution, (d) shear angle map indicates strong shear about the PIL of SP1 
   and SN1 with their rotation and shear motion of SP1, (e) Alpha 
   ($\alpha = (\nabla\times\mathbf{B})_z/B_z$) distribution with iso-contours 
   of LOS field indicates overal negative twist of the region (f) helicity flux 
   density map showing negative helicity distribution in the sunspots that is 
   consistent with negative ``alpha'', (g) Hinode continuum G-band image, and 
   (h) Hinode chromospheric Ca {\sc ii} image with the bright flare ribbons of 
   M2.2 at $14/17:20$UT around the PIL.}\label{roi1}
\end{figure}

The vector magnetic field map of the sub-region R1 on February
14/18:00UT is shown in Figure~\ref{roi1}(a) along with the map of
tracked horizontal velocities of magnetic fluxes in
Figure~\ref{roi1}(b). The magnetic transverse vectors are seen
nearly aligned with the polarity inversion line(PIL, thick dashed curve), 
implying a large shear along the SP1-SN1 interface. The map of tracked 
velocities shows the velocity vectors exhibiting spiral or vortical patterns 
in the penumbral region of SN1 as a fact of its rotation, and the velocity vectors 
aligned in the opposite direction to that of the magnetic field vectors. 

The new emerging positive flux SP1 rotated in the CCW direction,
along with a proper motion towards the main flare site R2. It is
noteworthy that when the footpoints were dragged along one
direction, the corresponding vector magnetic field pointed towards
the opposite direction to retain its footpoint connectivity. In the
process, the shear in magnetic structures built up. The shear motion
along the polarity inversion line (PIL) is an effective mechanism
for enhancing non-potentiality of magnetic fields. This can be
identified as the alignment of transverse magnetic field vectors
along the PIL \citep{ambastha1993}. Clearly, shear (Equation~\ref{Eq_ShD}) distribution 
map shows intense shear about the PIL with the rotation of SN1 and SP1 and proper 
motion of SN1 as well. This process indeed enhanced the electric currents along PIL 
as is evident from the the distribution of calculated vertical current $J_z$ in 
frame (c) of Figure~\ref{roi1}. Following \citet{zhangh2001}, the $J_z$ is decomposed to 
current of chirality $J_z^{\rm ch}$ and current of heterogeneity $J_z^{\rm h}$, and 
examined their trend. The average ratio $J_z^{\rm ch}$/$J_z^{\rm h}$ through out the 
time is deduced to be 1.11 implying the region is chirally dominated and is approximately 
consistent with force-free equilibrium. Interestingly, the shear current i.e. $J_z^{\rm h}$
follows similar increasing trend with $J_z^{\rm ch}$ having correlation coefficient of 0.97, 
inferring that the rotational motion leads to increase both twist as well as shear 
contributing to total current $J_z$. Moreover, the force free parameter $\alpha=J_z/B_z$ 
is having negative distribution (of order $10^{-6}$ m$^{-1}$) explaining the 
chirality with the counter rotation of the constituent sunspots in R1. From the 
AIA observations of R1, and also as confirmed by STEREO, it is observed that 
the intermittent mass expulsions turned into fast CMEs in this sub-region.

A map of helicity flux density computed from horizontal flux motions is shown in
Figure~\ref{roi1}(d). Through out the time interval, the helicity flux distribution is 
homogeneous with negative (dark) sign and implies left handed sense of chirality, 
further conforming to the observed CCW rotation of SP1 and SN1. The bipolar flux system 
having such counter-clock (SP1) and counter-clock (SN1) rotations forms sigmoidal 
patters which are efficient energy storage mechanisms to account for the
flares/CMEs \citep{canfield1999}. This is also evidenced by the high
resolution Hinode G-band continuum image showing filamentary
structures (frame (e)) that are consistent with the rotational motion
of SP1 and SN1. The corresponding Hinode chromospheric Ca {\sc ii}
image (frame (f)) shows the flare ribbons along the PIL and overlying
bright flare loops across the PIL (dashed line) associated with an
M-class event of February 14. These observations suggest to infer
that the rotational motion of SN1 played a significant role in
creating favorable conditions for the eruptions.

The vector magnetic map of sub-region R2 is shown in 
Figure~\ref{roi2} at the time of X2.2 flare with a smoothed polarity 
inversion line (thick dashed curve) separating positive and negative LOS 
magnetic field. This sub-region gave rise to several flares (listed in Table~\ref{TabFlCm}) 
of varying magnitudes including the energetic X2.2 flare along with its
associated CME. Importantly, a major portion of the flux (SP2 and SN2) is 
relatively associated to shearing motion, compared to flux within rotating 
SP2 itself, therefore this region R2 is shear motion dominated. As in R1, 
the alignment of $B_t$ vectors with PIL in this sub-region also showed strongly 
sheared magnetic fields. The tracked horizontal velocities (cf., \citealt{schuck2006}) displayed
strong shearing motion of SP2 along PIL. These motions led to the
stressing of field lines and possible flux cancellation as SP2 and
SN2 possessed opposite fluxes. The distribution of vertical currents
($J_z$), another indicator of large non-potentiality of magnetic
structures, too showed strong currents around the PIL in R2 in the form 
of J-shaped ribbons. High order shear is polarized along the PIL probably due 
to continuous shear motion of SP2 and rotation. This sub-region is roughly 
consistent with force-free equilibrium as the average ratio over the time  
$J_z^{\rm ch}$/$J_z^{\rm h}$ =1.29 indicating dominant chirality associated current. Nevertheless 
there exist dominant shear motion with magnetic shear and gradients, $J_z^{\rm h}$ does not show 
increasing trend, but indeed there do observed in $J_z^{\rm ch}$ as that in total current $J_z$. 
It means that both rotational as well as proper motions are contributing to $J_z^{\rm ch}$ the 
component that is parallel to to magnetic field. Therefore, the boundary motions in the evolution of the system
that is in force-free equilibrium retains it to be in same force-free equilibrium. Distribution of helicity flux 
is inhomogeneous having mixed polarities and especially during the peak times of some 
flare events like M6.6 and X2.2, we noticed the negative helicity flux in 
the existing system of positive helicity flux about the PIL. Injection of such opposite 
signed helicity is indicative of imminent transient activity. Whereas twisted penumbral 
fibril structure is seen in the Hinode G-band image on the photosphere, while, the bright  
chromospheric ribbons from X2.2 flare lying on either side of the PIL are seen
in the Hinode Ca {\sc ii} image.

\begin{figure}
  \centering
  \includegraphics[width=.486\textwidth,clip=,bb=15 3 372 783]{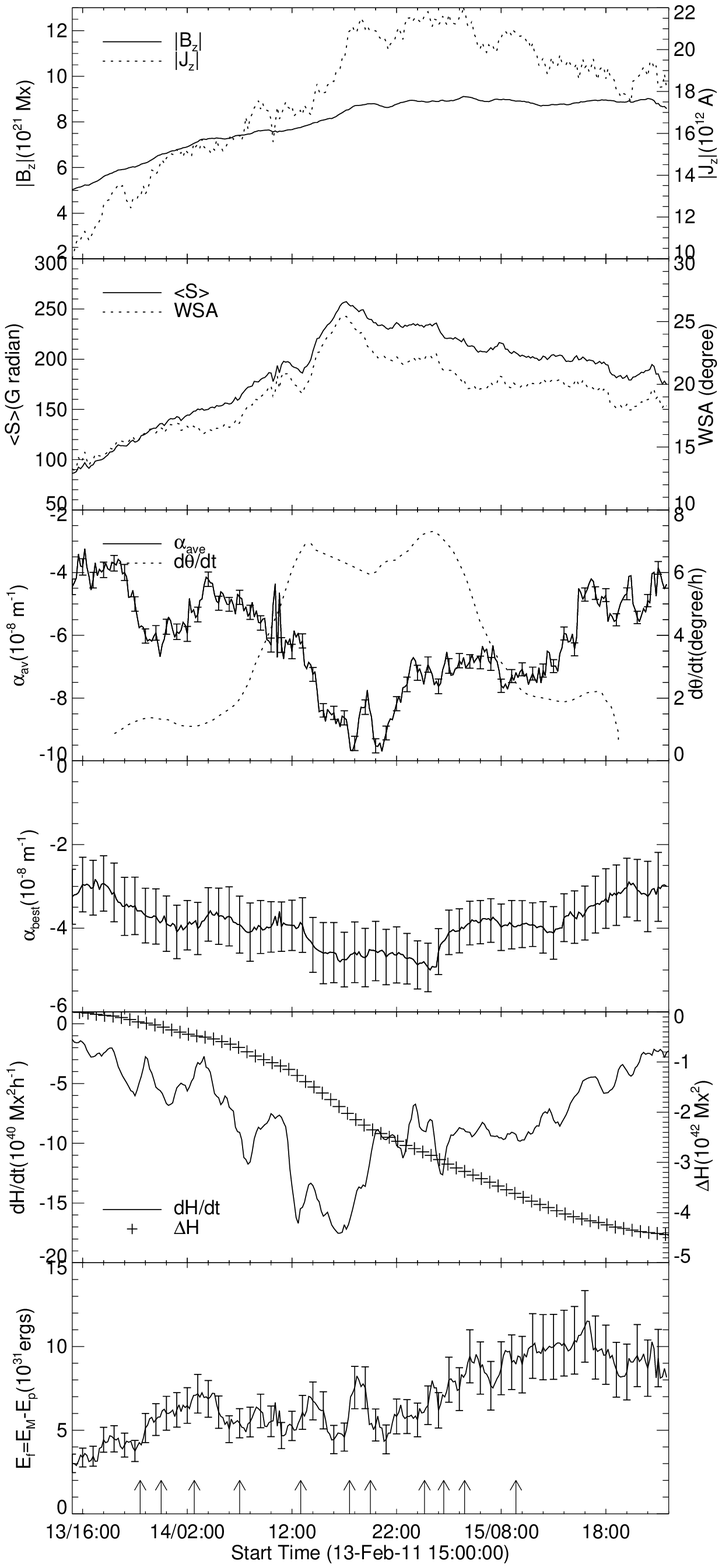}
  \includegraphics[width=.49\textwidth,clip=,bb=15 3 375 783]{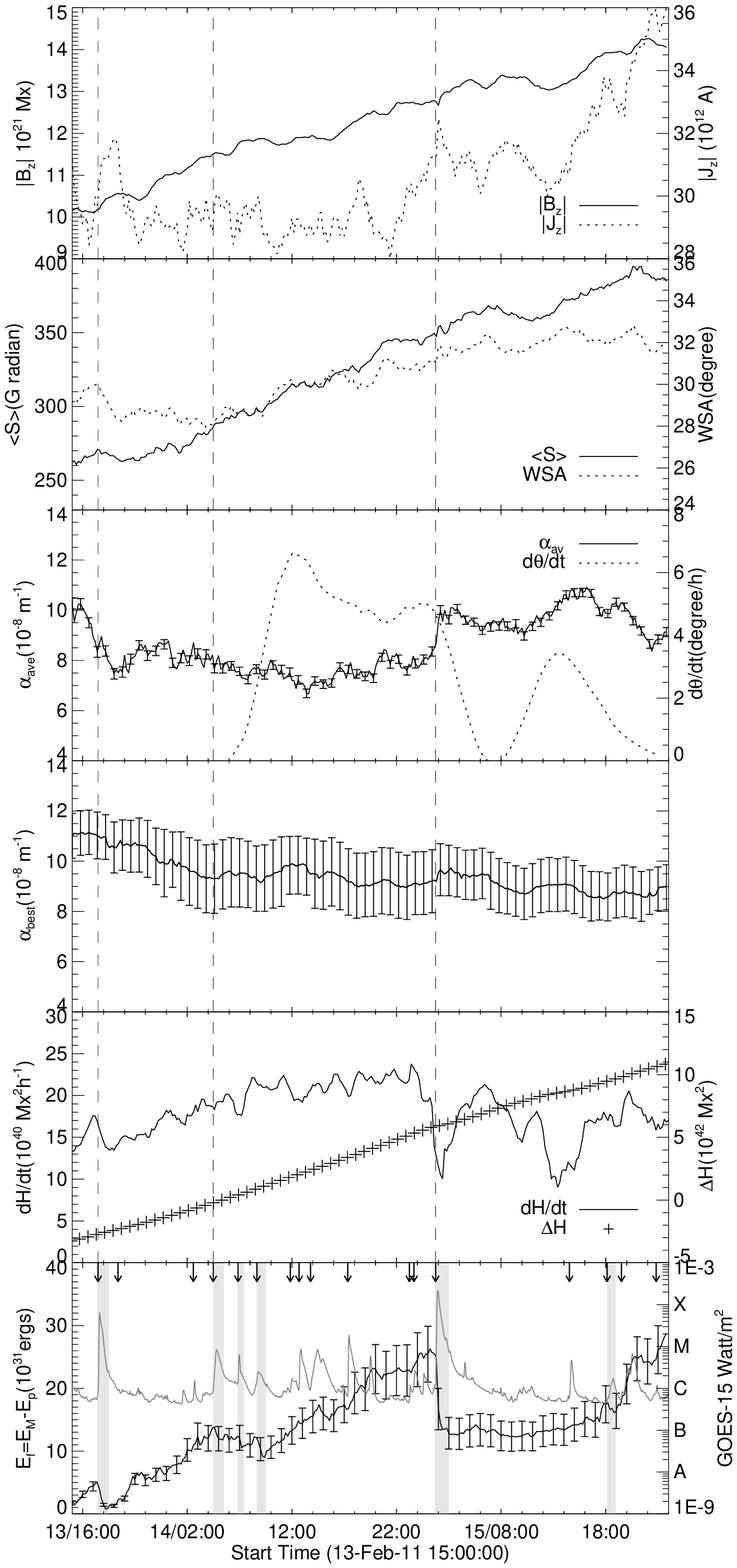}
   \caption{Temporal evolution of various physical parameters characterizing the 
   non-potentiality in sub-regions R1 (left column) and R2 (right column). Arrows 
   in the lower-most panel of the left column mark the CME timings while those in 
   the right column correspond to the GOES-15 flares. The three major flares are 
   marked by dashed vertical lines in all panels on the right column, for reference, 
   while the shaded vertical bars indicate the release of free-energy in step with 
   the onset of flares. Note that all the twist parameters $\alpha_{\rm av}$, 
   $\alpha_{\rm best}$, average shear ($<S>$), and dH/dt has clear correspondence 
   with the rotational profile of the sunspot SN1 implying that sunspot rotation has 
   direct role in increasing the non-potentiality in sub-region R1.}\label{plot_params}
\end{figure}

Temporal profiles of the deduced physical parameters for the
sub-region R1 are plotted in Figure~\ref{plot_params} (left column).
The total absolute flux doubled from $5\times10^{21}$Mx to
$9\times10^{21}$Mx from February 13 to February 15. Thereafter, it
showed no significant increase. This increase of flux is gradual which is 
contributed by emergence of SP1 and gradual evolution of SN1 in a proportion of 
approximately 1:3. The absolute vertical current also increased along with
the flux till February 15, and decreased afterwards. There was a
pronounced increase in $|J_z|$, from 2.0 to 2.6$\times10^{12}$A,
after 14/14:00UT. The average of shear ($<S>$) and weighted shear angle(WSA) are having 
similar trend, and explains the $J_z$ temporal trend. This might be 
related to the increasing rotation rate of the sunspot, thereby increasing the 
shear of horizontal magnetic fields by dragging the foot-points accommodating 
large field gradients about the PIL.

The sign and magnitude of $\alpha_{\rm av}$ and $\alpha_{\rm best}$
are found to be consistent as both the parameters are equivalent
proxies representing the twist. The larger error bars in
$\alpha_{\rm best}$ arise due to the $150$G error in the transverse
field given to propagate in the fitting procedure (See Eq.
\ref{Eq_Err_Alpav}). The negative sign indicates the left handed
twist or chirality. This agrees well with the physically observed
CCW rotation of the sub-region consisting of the sunspots SN1 and
SP1. The twist proxies, $\alpha_{\rm av}$ and $\alpha_{\rm best}$,
showed an increasing trend (in magnitude) with the rotation profile
of SP1 (see Figure~\ref{FigRot}) further corroborating the
increase in non-potentiality, as also evident from the absolute
vertical current $|J_z|$. A similar trend also reflected from the
temporal profile of helicity injection rate $dH/dt$ calculated from
the physically derived horizontal velocity field of the flux
motions. The distribution of helicity flux is homogeneous with negative 
sign over this region R1 and the helicity injection rate increased 
(in magnitude) with the rotation rate, reaching a maximum of 
$-17.52\times10^{40}$Mx$^2$h$^{-1}$. Thereafter it decreased as did the 
rotation rate. As corona can not accommodate any amount of helicity that 
is being accumulated continuously, it would try to expel to still outer 
atmosphere in the form of CMEs so that the system to be stable, 
therefore explaining the CMEs observed from region R1 of this AR. A 
total helicity accumulation of $-4.44\times10^{42}$Mx$^2$ was 
estimated in the sub-region R1 during this period. Since major part of the flux(SN1) 
is associated to rotational motion, this accumulated helicity is mostly 
contributed by dominant rotational motion in this sub-region.
 
The three parameters are different representations of twist or
magnetic complexity except the manner in which they are estimated.
Increased rotation rate of SN1 seems to increase the twist so does
the increasing trend of these parameters. The strong mass expulsions
observed in AIA/304\AA~at 12:50, 17:30, 19:20UT on February 14,
turning into CMEs, further strengthen this evidence. This
correspondence of the twist parameters $\alpha_{\rm av}$, $\alpha_{\rm best}$,$<S>$, WSA, 
and $dH/dt$ with the rotational profile of SN1 suggests the
predominant role of rotation in the observed expulsions/CMEs (refer
to panels a1 and b1 in Figure~\ref{aia_plot}). 

Thus we are able to identify the correspondence of the measured
twist parameters with the observed rotation of the sunspot in
sub-region R1. However, the free-energy deduced for R1 did not show
any obvious association with the observed CMEs. We believe that the
broad peaks in free-energy profile seen during the CMEs (12:50,
17:30, 19:20UT) might have some relation but a rapid release of
energy is not obvious. The maximum mean value of free-energy was
estimated around $10^{32}$ ergs; adequate to account for the large
eruptions.

Temporal profiles of the deduced physical parameters for the
sub-region R2 are plotted in Figure~\ref{plot_params} (right
column). Absolute flux for this sub-region was much larger in
magnitude as compared to that in R1. It increased monotonically from
$10\times10^{21}$Mx to $14\times10^{21}$Mx. The magnitude of
absolute current $J_z$  too was correspondingly larger and showed
appreciable variations associated with local evolution. Peaks in the
electric current profile are noticeable at the times of large flare
events, i.e., M6.6 and X2.2. The currents decreased thereafter,
implying the build up currents in the preflare phase leads to relaxation of the 
magnetic field structure by releasing energy during flares. These peaks does not 
reflect in the $<S>$ and WSA profile because of averaging property of these parameters. 
Note that WSA is higher by about 7\arcdeg~compared to that of R1 implying high level 
of non-potentiality over the region. Both the 
twist proxies, $\alpha_{\rm av}$ and $\alpha_{\rm best}$, agreed in
sign and magnitude. Unlike in R1, the positive sign of these
parameters represents a positive or right handed twist in R2. A
relatively large average value of $\alpha_{\rm av}$ at
$8.5\times10^{-8}$m$^{-1}$ is suggestive of a stronger twist present
in this sub-region. There seems no corresponding trend of any of these parameters 
with the rotational profile of SP2 probably complexity in the flux system is not 
only due to twisting but also shearing of fluxes.

Unlike R1, the sub-region R2 showed positive helicity injection at
a much larger rate, with a maximum of $29.11\times10^{40}$Mx$^2$h$^{-1}$.
The accumulated helicity too was comparatively larger. A total
helicity of $10.27\times10^{42}$Mx$^2$ was injected during the 60h
period by the observed flux motions and as a result the coronal helicity is likely to 
be positive. Within R2, shearing of flux system between SN2 and SP2 is 
dominated over twisting of single flux system of SP2, therefore it is likely that 
the accumulated helicity is mostly contributed by the shear motion. This 
observance is consistent with reports of \citet{liuj2006} in AR 10488 having 
shearing and twisting phases. They claimed that shear motion between two 
different flux systems indeed the major contributor of helicity injection 
in that AR.

At the epochs of the M6.6 and X2.2 flares, sharp dips in helicity rate occurred 
corresponding to the 
injection of opposite, negative helicity. Existence of such both 
signs of helicity flux in a single domain is supposed to be sign of triggering 
transient flare events \citep{linton2001, kusano2004}. However, such an observational 
evidence of negative or opposite helicity flux may not be real change due to true 
helicity transfer, it could be flare related transient change effecting the magnetic 
field measurements. The observed distribution of negative helicity flux coincides 
spatially and temporally to the reports of possible flare effects on Doppler 
and magnetic measurements in X2.2 flare \citep{maurya2012}. A specific 
study focusing the helicity injection and its behavioral change during 
flares and CMEs from this AR 11158 in presented in \citet{vemareddy2012b}. They 
showed the change of helicity flux signal due to injection of opposite helicity 
could be a true change unless there is no impulsive effect of the of the flare 
on magnetic field measurements. In particular, the localized helicity flux distribution 
is affected by flare-related transient effects during these M6.6 and X2.2 flare 
cases having sudden dips in the helicity flux profile as drawn in the figure. 

The temporal profile of free-energy is plotted in the bottom frame
of Figure~\ref{plot_params}, along with the GOES X-ray flux. The
onset times of flares from the sub-region R2 are marked in this
frame. The error bars represent standard deviation of free-energy
obtained by changing the origin of coordinate system as explained
before, representing the departure from force-freeness and flux
balance. It is to note that the AR, as a whole, nearly fulfills the
conditions for force-freeness and flux balance ($\leqslant0.05$).
Therefore, the constructed potential field from the extrapolation is
not far from these conditions. The integration was carried out over
the sub-regions of interest, i.e., R1 and R2, using
eq~\ref{Eq_FreeEne}. Although the force-free condition was nearly
satisfied, a part of field lines from R2 were seen connected to R1
from the coronal AIA images. Hence, the condition for flux balance
was not fully obeyed, posing some problem for the applicability of
the virial-theorem. Therefore, the derived results are not
conclusive enough unless the mean value of free-energy is much
larger than the error bars.

The most significant finding of the free-energy plot is the step-down 
decrease in free energy at the onset of the flares (indicated by the
GOES X-ray profile). This provides an unambiguous evidence of energy
release during the flares. Magnetic energy is believed to be stored
by flux emergence and/or shear motions and is reflected in the form
of electric currents. When an instability occurs in the field
structure, the excess or free energy is released in the form of
flares and CMEs. The flare events having such a correspondence are
shown by vertical shaded bars.

As the magnetic fluxes evolved and moved out of the sub-area R2, the
fields departed increasingly from the flux-balance condition.
Therefore, the free energy derived after the X2.2 flare are
increasingly unmeaningful as evident from the large error bars. We
estimated that the flux imbalance steadily increased from 5\% on
February 13 to 16\% on February 15. However, the mean value of free
energy showed a clear step down decrease at the onset of the X2.2
flare, indicating an appreciable release of energy sufficient to
account for this event. Therefore, the relatively large error bars
notwithstanding, the variation of free-energy appears to be
convincing at the time of this energetic X-class flare.

The changes in potential and total energy estimated during the
period from the onset to post-flare phase are given in
Table~\ref{FlrEne} for some flares where we found the step-down
decrease in the energy. The difference in available free-energy
between the post-flare and the flare onset time giving the amount of
energy released during the main flares, are described in the
following.
\begin{table}[htbp]
  \centering
  \caption{Estimated total and potential energies during flares}
    \begin{tabular}{rrrrr}
    \addlinespace
    \toprule
    Flare & \multicolumn{2}{c}{Total Energy($E_T$ $10^{32}$ergs)} & \multicolumn{2}{c}{Potential Energy($E_P$ $10^{32}$ergs)} \\
    \midrule
          & \multicolumn{1}{c}{Before Flare} & \multicolumn{1}{c}{After Flare} & \multicolumn{1}{c}{Before flare} & \multicolumn{1}{c}{After flare} \\
    M6.6  & $1.11\pm0.39$ & $0.69\pm0.25$   & $0.61\pm0.22$  &  $0.63\pm0.24$   \\
    X2.2  & $4.92\pm1.58$ & $3.67\pm1.20$   & $2.43\pm0.78$  &  $2.47\pm0.79$   \\
    C8.3  & $3.35\pm1.06$ & $3.16\pm1.00$   & $1.99\pm0.63$  &  $1.97\pm0.63$  \\
    C6.6  & $3.25\pm1.03$ & $3.20\pm1.01$   & $2.02\pm0.64$  &  $2.11\pm0.67$   \\
    C1.8  & $3.32\pm1.05$ & $2.97\pm0.94$   & $2.11\pm0.67$  &  $2.07\pm0.66$   \\
    C1.7  & $2.27\pm0.79$ & $2.29\pm0.81$   & $2.88\pm0.82$  &  $2.56\pm0.08$   \\
    \bottomrule
    \end{tabular}%
  \label{FlrEne}%
\end{table}%

\noindent The X2.2 flare of February 15/01:44UT: The total energy
$E_T \approx [4.92\pm1.58]\times10^{32}$ erg was estimated on February
15/01:48UT. It decreased to $[3.67\pm1.20]\times10^{32}$ erg after
the impulsive phase of the flare, i.e., at 02:12UT. The ratio
$E_T/E_P$ of the total energy with the energy of the respective
potential state correspond to 2.01 and 1.46 at these two time
instants. During this short 24 minute period, the excess energy
released was estimated as $[1.05\pm1.04]\times10^{32}$erg,
sufficient to account for an X-class flare.

\noindent The M6.6 flare of February 13/17:28UT:  The ratio
$E_T/E_P$ of 1.6 at the onset time of the flare at 17:24UT reduced
to 1.08 at 18:12UT, i.e., after the flare. This implies the release
of free-energy of $[4.3\pm1.7]\times10^{31}$erg, i.e., consistent
with the magnitude of the flare.

Similarly, the free-energies released for the C-class events, listed
in the Table~\ref{TabFlCm}, shown by shaded bars and were estimated as
$(1.8\pm5.9)\times10^{31}$, $(1.6\pm5.4)\times10^{31}$,
$(3.7\pm5.2)\times10^{31}$ and $(1.1\pm4.5)\times10^{31}$erg. These
estimated values are not statistically significant as the mean
values are smaller than the errors by a factor of two. Because the
magnetic energy change is purely due to the change in the boundary
conditions, it is expected that even minute changes in the magnetic
field corresponding to a C-class flare would reflect in the mean
values retaining error bars without much change.

%%%%%%%%%%%%%%%%%%%%%%%%%%%%%%%%%%%
%%    CONCLUSIONS                %%
%%%%%%%%%%%%%%%%%%%%%%%%%%%%%%%%%%%
\section{Summary and Conclusions}
\label{SumConc} This study suggests that the rotational motions of major
sunspots in NOAA AR 11158 were not only related to the transport of magnetic
energy and complexity from the low atmosphere to the corona, but also appear to
have played a major role in the onset of flares and CMEs. We have attempted to
infer the relationship between the observed flux motions, variations in derived
physical parameters in association with the dynamical transients, viz., flares
and CMEs. A careful analysis of the flux motions leads us to better understand
the dynamical nature of the photosphere at the interface between the convection
zone where basic dynamo originates and the tenuous corona.

The observations presented here provide a direct
evidence for the energization of the solar corona by the emergence
and/or dynamics of magnetic flux-tubes that appear as rotating
sunspots. This phenomenon is important for understanding how the
solar atmosphere attains the conditions necessary for the release of
energy and helicity observed in solar flares and CMEs.
For a detailed study of various physical parameters characterizing
non-potentiality, we selected two sub-regions, R1 and R2, of AR NOAA
11158, which consisted of rotating sunspots. In R1 with the rotating
sunspot SN1, the estimated rotation rate is found to have a good
correspondence with twist parameters $\alpha_{\rm av}$ and
$\alpha_{\rm best}$, $<S>$, and dH/dt. It showed that the intrinsic rotation of the
sunspot increased the overall twist in sub-region R1. Major CMEs in
R1 occurred on 14 February at 12:50, 17:30, 19:20UT, which coincided
well with the timings of large helicity injection rates and twist
parameters. As the sunspot's rotation rate reached the maximum, the
fluxes moved with maximum velocities to provide larger complexity in
the connectivity. The calculated free-energies, however, did not
show good correspondence with the observed CME events, except that 
broad peaks were observed during these CMEs. This is largely due to
violation of flux-balance condition(80\%). It is to note that the
occurrence of CMEs is also attributed to the new flux emergence
\citep{martin1985,zhang2008}. However, the absolute flux profile of
R1 does not show such sudden emergence of flux related to emergence of SP1.

\begin{figure}
  \centering
  \includegraphics[width=1.0\textwidth,clip=,bb=61 37 560 306]{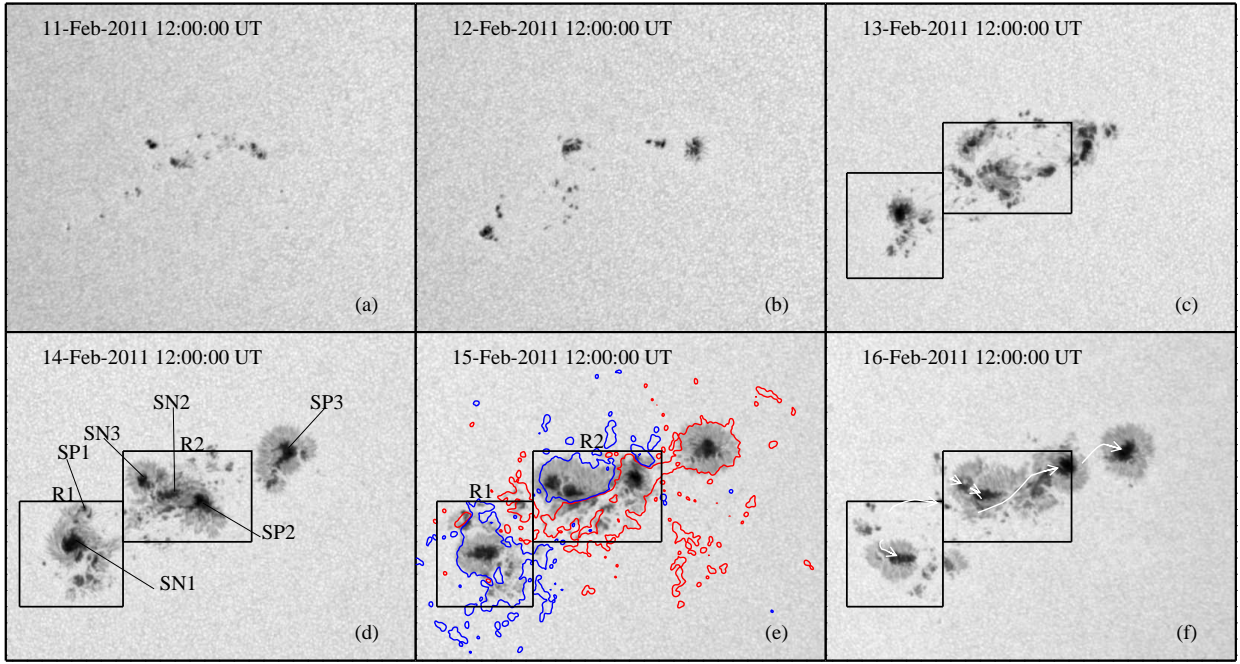}
  \caption{ HMI intensity images showing the evolution of AR NOAA 11158 during 
  six day time period. The main sunspots are labeled by SP/N* in (d) along 
  with the LOS magnetic field contours overlaid in red (blue) color at 150(-150)G 
  levels in panel(also in subsequent figures unless specified) (e). Proper 
  motions of individual sunspots are traced along the arrowed curves as in 
  panel (f). The two rectangular boxes mark the selected sub-regions R1 and 
  R2 for further study.}
  \label{IntMos}
\end{figure}

\begin{figure}
  \centering
  \includegraphics[width=.65\textwidth,clip=,bb=66 41 560 773]{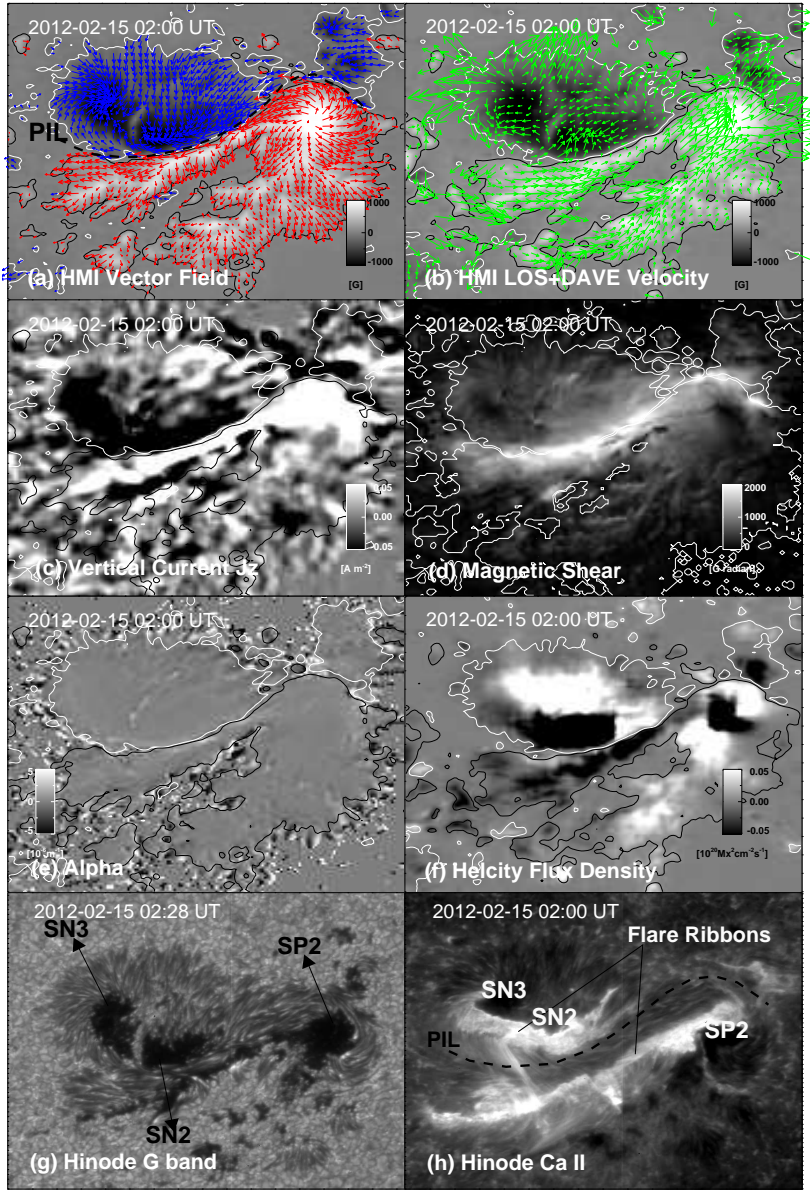}
   \caption{Sub-region R2 on February 15/02:00UT: (a) transverse magnetic field 
   ($B_t$) vectors overlaid on LOS magnetic field with PIL separating positive 
   and negative polarities, (b) tracked horizontal velocity vectors overplotted 
   on LOS magnetic field, (c) Vertical current distribution ($J_z$) showing intense 
   currents accumulated on either side of PIL with the shear motion of SP2, (d) Shear 
   angle map also implies strong shear polarized about the PIL with the continuous 
   shear motion of SP2, (e) Distribution of twist parameter ``alpha'' indicates 
   positive twist of the region, (f) helicity flux density map with intense negative 
   helicity flux about PIL probably due to flare-effects on magnetic field 
   measurements, (g) Hinode continuum G-band, and (h) Hinode chromospheric Ca {\sc ii} 
   image showing the flare ribbons of $X2.2$ around the PIL (dashed curve).}\label{roi2}
\end{figure}

No clear relationship of the rotation rate was found with the twist
parameters and dH/dt for the sunspot SP2 in R2. Shear motion of SP2
(cf., Figure~\ref{IntMos}f) seems to be the dominant factor, which
obliterates the rotational effect of SP2 on the twist parameters and
helicity rate. This shear motion caused the storage of magnetic
energy by stressing the field lines. The intermittent release of
energy during the observed flares is discernible. A total
free-energy of $(1.05\pm1.04)\times10^{32}$erg is released during
the X2.2 flare. This is three times larger than the derived energy
from coronal magnetic field extrapolations by \citet{xudong2012}
over the entire AR. Spectral line-profile reversals occurred during
the impulsive phase of this flare affecting the magnetic and
velocity field measurements, as reported recently by
\citet{maurya2012}. As a result, the injected helicity rate showed
an impulsive negative peak by the sudden appearance of negative
helicity density indicating flare-related transient effect (cf., Figure~\ref{roi2}d).

The M6.6 flare of February 13/17:28UT was another large but less
impulsive event which released $(4.3\pm1.7)\times10^{31}$erg of
energy in the 48 minute period. Step down decrease of free-energy
was found not only during the two large flares, but also in some
flares of smaller magnitude. This reveals the storage and release of
energy in the flares occurring in sub-region R2. It is to mention
that we have not carried out the analysis of errors due to the
spectro-polarimetric as well as random noise and their effects on
the free-energy estimation, as there is no established method
available on the data set as yet. Moreover, the estimation of free-energy 
suffers by the departure from the validity conditions of virial theorem 
due to the selection of small sized areas of the sub-regions. 

The accumulated helicity contributed by the rotational motion of the sunspot
can also be obtained as                   %Zhang Hongqui 2008, solarphysics%
\begin{equation}
\Delta H=-\frac{1}{2\pi}\Delta \theta \Phi^{2},
\end{equation}
where $\Phi$ is the magnetic flux and $\Delta\theta$ is the total
rotation angle of the sunspot. Since the sunspot SN1 rotated by
$\approx 160\arcdeg$ during the 60h period of our study, and
assuming the average magnetic flux of this sunspot as
$3.5\times10^{21}$Mx, the accumulated helicity by rotation is
estimated as -$5.44\times10^{42}$Mx$^2$. This is consistent with the
derived value of -$4.44\times10^{42}$Mx$^2$ from the tracked velocities of 
the fluxes. 

Thus, this study demonstrates that both shear (dominated in R2) and rotational 
(dominated in R1) motions of the observed fluxes enhanced the magnetic
non-potentiality of the active region by injecting helicity. This
helped in increasing the free energy of the AR's magnetic field by
increasing the overall complexity leading to the conditions favorable for
the eruptions. The deduced physical parameters describing the level
of magnetic non-potentiality and the free energy showed a reasonably
good correspondence with the observed transient activity of the AR.
This study also provides a clear signatures of energy release during
the major, energetic transients of the AR.

\acknowledgements We thank an anonymous referee for the comments that 
helped us to improve the readability of the manuscript. The data have been used here 
courtesy of NASA/SDO and HMI science team. We thank the HMI team for 
making available the processed vector magnetic field data. The authors are 
grateful to Dr. Xudong Sun of Stanford University for his help in handling the
vector magnetogram data. This work utilizes the Ca {\sc ii} H and
Continuum data from the Solar Optical Telescope (SOT) on board Hinode.
%%%%%%%%%%%%%%%% BIBLIOGRAPHY AND ACKNOWLDGEMENTS %%%%%%%%%%%%%%%%%%%%%%%%%%%%%%
\bibliographystyle{apj}
%%\bibliography{ref_bulk}

\begin{thebibliography}{44}
\expandafter\ifx\csname natexlab\endcsname\relax\def\natexlab#1{#1}\fi

\bibitem[{{Ambastha} {et~al.}(1993){Ambastha}, {Hagyard}, \&
  {West}}]{ambastha1993}
{Ambastha}, A., {Hagyard}, M.~J., \& {West}, E.~A. 1993, \solphys, 148, 277

\bibitem[{{Bhatnagar}(1967)}]{batnagar1967}
{Bhatnagar}, A. 1967, \kob

\bibitem[{{Borrero} {et~al.}(2011){Borrero}, {Tomczyk}, {Kubo},
  {Socas-Navarro}, {Schou}, {Couvidat}, \& {Bogart}}]{borrero2011}
{Borrero}, J.~M., {Tomczyk}, S., {Kubo}, M., {Socas-Navarro}, H., {Schou}, J.,
  {Couvidat}, S., \& {Bogart}, R. 2011, \solphys, 273, 267

\bibitem[{{Brown} {et~al.}(2003){Brown}, {Nightingale}, {Alexander},
  {Schrijver}, {Metcalf}, {Shine}, {Title}, \& {Wolfson}}]{brown2003}
{Brown}, D.~S., {Nightingale}, R.~W., {Alexander}, D., {Schrijver}, C.~J.,
  {Metcalf}, T.~R., {Shine}, R.~A., {Title}, A.~M., \& {Wolfson}, C.~J. 2003,
  \solphys, 216, 79

\bibitem[{{Canfield} {et~al.}(1999){Canfield}, {Hudson}, \&
  {McKenzie}}]{canfield1999}
{Canfield}, R.~C., {Hudson}, H.~S., \& {McKenzie}, D.~E. 1999, \grl, 26, 627

\bibitem[{{Chandrasekhar}(1961)}]{chandrasekhar1961}
{Chandrasekhar}, S. 1961, {Hydrodynamic and hydromagnetic stability}, ed.
  {Chandrasekhar, S.}

\bibitem[{{Evershed}(1910)}]{evershed1910}
{Evershed}, J. 1910, \mnras, 70, 217

\bibitem[{{Freeland} \& {Handy}(1998)}]{freeland1998}
{Freeland}, S.~L., \& {Handy}, B.~N. 1998, \solphys, 182, 497

\bibitem[{{Gary}(1989)}]{gary1989}
{Gary}, G.~A. 1989, \apjs, 69, 323

\bibitem[{{Gary}(2001)}]{gary2001}
---. 2001, \solphys, 203, 71

\bibitem[{{Hagino} \& {Sakurai}(2004)}]{hagino2004}
{Hagino}, M., \& {Sakurai}, T. 2004, \pasj, 56, 831

\bibitem[{{Hiremath} {et~al.}(2005){Hiremath}, {Suryanarayana}, \&
  {Lovely}}]{hiremath2005}
{Hiremath}, K.~M., {Suryanarayana}, G.~S., \& {Lovely}, M.~R. 2005, \aap, 437,
  297

\bibitem[{{Hoeksema} \& {et al.}(2012)}]{hoeksema2012}
{Hoeksema}, J.~T., \& {et al.} 2012, \solphys, To be Submitted

\bibitem[{{Howard} {et~al.}(1990){Howard}, {Harvey}, \& {Forgach}}]{howard1990}
{Howard}, R.~F., {Harvey}, J.~W., \& {Forgach}, S. 1990, \solphys, 130, 295

\bibitem[{{Jiang} {et~al.}(2012){Jiang}, {Zheng}, {Yang}, {Hong}, {Yi}, \&
  {Yang}}]{jiang2012}
{Jiang}, Y., {Zheng}, R., {Yang}, J., {Hong}, J., {Yi}, B., \& {Yang}, D. 2012,
  \apj, 744, 50

\bibitem[{{Kusano} {et~al.}(2004){Kusano}, {Maeshiro}, {Yokoyama}, \&
  {Sakurai}}]{kusano2004}
{Kusano}, K., {Maeshiro}, T., {Yokoyama}, T., \& {Sakurai}, T. 2004, \apj, 610,
  537

\bibitem[{{Leka} \& {Barnes}(2003)}]{leka2003a}
{Leka}, K.~D., \& {Barnes}, G. 2003, \apj, 595, 1277

\bibitem[{{Leka} {et~al.}(2009){Leka}, {Barnes}, {Crouch}, {Metcalf}, {Gary},
  {Jing}, \& {Liu}}]{leka2009}
{Leka}, K.~D., {Barnes}, G., {Crouch}, A.~D., {Metcalf}, T.~R., {Gary}, G.~A.,
  {Jing}, J., \& {Liu}, Y. 2009, \solphys, 260, 83

\bibitem[{{Lemen} {et~al.}(2012){Lemen}, {Title}, {Akin}, {Boerner}, \& {et
  al}}]{lemen2012}
{Lemen}, J.~R., {Title}, A.~M., {Akin}, D.~J., {Boerner}, P.~F., \& {et al}.
  2012, \solphys, 275, 17

\bibitem[{{Linton} {et~al.}(2001){Linton}, {Dahlburg}, \&
  {Antiochos}}]{linton2001}
{Linton}, M.~G., {Dahlburg}, R.~B., \& {Antiochos}, S.~K. 2001, \apj, 553, 905

\bibitem[{{Liu} \& {Zhang}(2006)}]{liuj2006}
{Liu}, J., \& {Zhang}, H. 2006, \solphys, 234, 21

\bibitem[{{Low}(1982)}]{low1982}
{Low}, B.~C. 1982, \solphys, 77, 43

\bibitem[{{Martin} {et~al.}(1985){Martin}, {Livi}, \& {Wang}}]{martin1985}
{Martin}, S.~F., {Livi}, S.~H.~B., \& {Wang}, J. 1985, Australian Journal of
  Physics, 38, 929

\bibitem[{{Maurya} {et~al.}(2012){Maurya}, {Vemareddy}, \&
  {Ambastha}}]{maurya2012}
{Maurya}, R.~A., {Vemareddy}, P., \& {Ambastha}, A. 2012, \apj, 747, 134

\bibitem[{{McIntosh}(1981)}]{mcintosh1981}
{McIntosh}, P.~S. 1981, in The Physics of Sunspots, ed. {L.~E.~Cram \&
  J.~H.~Thomas}, 7--54

\bibitem[{{Metcalf}(1994)}]{metcalf1994}
{Metcalf}, T.~R. 1994, \solphys, 155, 235

\bibitem[{{Metcalf} {et~al.}(1995){Metcalf}, {Jiao}, {McClymont}, {Canfield},
  \& {Uitenbroek}}]{metcalf1995}
{Metcalf}, T.~R., {Jiao}, L., {McClymont}, A.~N., {Canfield}, R.~C., \&
  {Uitenbroek}, H. 1995, \apj, 439, 474

\bibitem[{{Metcalf} {et~al.}(2005){Metcalf}, {Leka}, \& {Mickey}}]{metcalf2005}
{Metcalf}, T.~R., {Leka}, K.~D., \& {Mickey}, D.~L. 2005, \apjl, 623, L53

\bibitem[{{Molodensky}(1974)}]{molodensky1974}
{Molodensky}, M.~M. 1974, \solphys, 39, 393

\bibitem[{{Pariat} {et~al.}(2005){Pariat}, {D{\'e}moulin}, \&
  {Berger}}]{pariat2005}
{Pariat}, E., {D{\'e}moulin}, P., \& {Berger}, M.~A. 2005, \aap, 439, 1191

\bibitem[{{Pevtsov} {et~al.}(1994){Pevtsov}, {Canfield}, \&
  {Metcalf}}]{pevtsov1994}
{Pevtsov}, A.~A., {Canfield}, R.~C., \& {Metcalf}, T.~R. 1994, \apjl, 425, L117

\bibitem[{{Schou} {et~al.}(2012){Schou}, {Scherrer}, {Bush}, {Wachter}, \& {et
  al}}]{schou2012}
{Schou}, J., {Scherrer}, P.~H., {Bush}, R.~I., {Wachter}, R., \& {et al}. 2012,
  \solphys, 275, 229

\bibitem[{{Schuck}(2006)}]{schuck2006}
{Schuck}, P.~W. 2006, \apj, 646, 1358

\bibitem[{{Sun} {et~al.}(2012){Sun}, {Hoeksema}, {Liu}, {Wiegelmann},
  {Hayashi}, {Chen}, \& {Thalmann}}]{xudong2012}
{Sun}, X., {Hoeksema}, J.~T., {Liu}, Y., {Wiegelmann}, T., {Hayashi}, K.,
  {Chen}, Q., \& {Thalmann}, J. 2012, \apj, 748, 77

\bibitem[{{Tian} \& {Alexander}(2006)}]{tian2006}
{Tian}, L., \& {Alexander}, D. 2006, \solphys, 233, 29

\bibitem[{{Tian} {et~al.}(2008){Tian}, {Alexander}, \&
  {Nightingale}}]{tian2008}
{Tian}, L., {Alexander}, D., \& {Nightingale}, R. 2008, \apj, 684, 747

\bibitem[{{Tsuneta} {et~al.}(2008){Tsuneta}, {Ichimoto}, {Katsukawa}, {Nagata},
  \& {et al}}]{tsuneta2008}
{Tsuneta}, S., {Ichimoto}, K., {Katsukawa}, Y., {Nagata}, S., \& {et al}. 2008,
  \solphys, 249, 167

\bibitem[{{Vemareddy} {et~al.}(2012){Vemareddy}, {Ambastha}, {Maurya}, \&
  {Chae}}]{vemareddy2012b}
{Vemareddy}, P., {Ambastha}, A., {Maurya}, R.~A., \& {Chae}, J. 2012, ArXiv
  e-prints

\bibitem[{{Wang} {et~al.}(1994){Wang}, {Ewell}, {Zirin}, \& {Ai}}]{wangh1994}
{Wang}, H., {Ewell}, Jr., M.~W., {Zirin}, H., \& {Ai}, G. 1994, \apj, 424, 436

\bibitem[{{Yan} {et~al.}(2008){Yan}, {Qu}, \& {Kong}}]{yan2008b}
{Yan}, X.-L., {Qu}, Z.-Q., \& {Kong}, D.-F. 2008, \mnras, 391, 1887

\bibitem[{{Yan} {et~al.}(2009){Yan}, {Qu}, {Xu}, {Xue}, \& {Kong}}]{yan2009}
{Yan}, X.-L., {Qu}, Z.-Q., {Xu}, C.-L., {Xue}, Z.-K., \& {Kong}, D.-F. 2009,
  Research in Astronomy and Astrophysics, 9, 596

\bibitem[{{Zhang}(2001)}]{zhangh2001}
{Zhang}, H. 2001, \apjl, 557, L71

\bibitem[{{Zhang} {et~al.}(2007){Zhang}, {Li}, \& {Song}}]{zhang2007}
{Zhang}, J., {Li}, L., \& {Song}, Q. 2007, \apjl, 662, L35

\bibitem[{{Zhang} {et~al.}(2008){Zhang}, {Liu}, \& {Zhang}}]{zhang2008}
{Zhang}, Y., {Liu}, J., \& {Zhang}, H. 2008, \solphys, 247, 39

\end{thebibliography}

%%%%%%%%%%%%%%%%%%%%%%%%%%%%%%%%%%%%%%%%%%%%%%%%
%%%%%%%%%%%%%%%%%%FIGURES%%%%%%%%%%%%%%%%%%%%%%%
%%%%%%%%%%%%%%%%%%%%%%%%%%%%%%%%%%%%%%%%%%%%%%%%

\end{document}